\documentclass[11pt]{article}
\usepackage{amsmath,amssymb,amsfonts}
\usepackage{mathbbol,bm,bbm}
\usepackage{hyperref}
\usepackage{graphicx}

\newcommand{\Cx}{{\mathbb C}}

\newcommand{\Rl}{{\mathbb R}}

\newcommand{\idty}{\mathbb{1}}
\DeclareMathOperator{\id}{id}

\DeclareMathOperator*{\tr}{Tr}
\newcommand{\<}{\langle}
\renewcommand{\>}{\rangle}

\providecommand{\norm}[1]{\Vert #1 \Vert}

\renewcommand{\c}[1]{\mathcal{#1}}

\newcommand{\s}[1]{\mathsf{#1}}
\renewcommand{\r}[1]{\mathrm{#1}}
\DeclareMathOperator*{\loplus}{\mbox{\Large\mbox{$\oplus$}}}

\usepackage{latexsym}

\begin{document}
\begin{center}
{\LARGE Davies maps for qubits and qutrits} 
 \\[12pt]
Wojciech~Roga$^1$, Mark~Fannes$^2$ and Karol~{\.Z}yczkowski$^{1,3}$
\end{center}

\medskip
\noindent
$^1$ Instytut Fizyki im.~Smoluchowskiego,
Uniwersytet Jagiello{\'n}ski,
PL-30-059 Krak{\'o}w, Poland \\
$^2$ Instituut voor Theoretische Fysica,
Universiteit Leuven, B-3001 Leuven, Belgium \\
$^3$Centrum Fizyki Teoretycznej, Polska Akademia Nauk,
PL-02-668 Warszawa, Poland

\medskip
\noindent
Email: \texttt{<wojciech.roga@uj.edu.pl>}, \texttt{<mark.fannes@fys.kuleuven.be>}, and \texttt{<karol@tatry.if.uj.edu.pl>} 

\bigskip
\noindent
\textbf{Abstract:}

We investigate the dynamics of an $N$-level quantum system weakly coupled to a thermal reservoir. For any fixed temperature of the bath there exists a natural reference state: the equilibrium state of the system. Among all quantum operations on the system one distinguishes Davies maps, they preserve the equilibrium state, satisfy the detailed balance condition and belong to a semi-group. A complete characterization of the three dimensional set of qubit Davies maps is given. We analyze these maps and find their minimum output entropy. A characterization of Davies maps for qutrits is also provided.

\medskip
\noindent
PACS: 
02.10.Ud (Mathematical methods in physics, Linear algebra), 
03.67.-a (Quantum mechanics, field theories, and special relativity, Quantum information), 03.65.Yz (Decoherence; open systems; quantum statistical methods)

\section{Introduction}

The quantum dynamics of simple model systems is a subject of considerable research interest in view of applications in quantum information processing~\cite{Ve06,Pe08}. Any discrete dynamics of an $N$-level system, which can be realized in a physical experiment, can be described by a \emph{quantum operation}, also called stochastic map. This is an affine transformation of the set of quantum states of size $N$ that is completely positive, see e.g.~\cite{AF01,BZ06}.

Within the set of quantum operations one distinguishes the subset of \emph{bistochastic maps,} they additionally satisfy the unitality condition: the uniform state is preserved. On one hand the set of bistochastic maps is easier to study and these maps enjoy several particular properties, for instance, the minimal output entropy is known to be additive for one-qubit bistochastic maps~\cite{Ki02}. On the other hand, maps that are not bistochastic do play a crucial role in the description of several physical problems.

Even in the case $N = 2$ the set of one-qubit operations is $12$ dimensional and it grows as $N^4$ for large $N$. Thus it is legitimate to ask, which of these maps could be of any special importance. To suggest any concrete class of maps one needs to make some physical assumptions and to investigate their implications.

\begin{figure}[ht]
\centering
\scalebox{.45}{\includegraphics{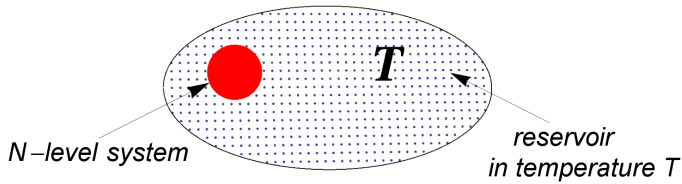}}
\caption{An $N$-level system weakly coupled to a reservoir at temperature $T$.}
\label{fig:sketch}
\end{figure}

The main aim of this work is to select a physically motivated class of quantum operations, which includes also non-unital maps, and to study their properties. In theoretical investigations one often deals with isolated quantum systems, which are entirely separated from the environment. Such an assumption is not realistic, so we are going to assume that the $N$-level system under investigation is weakly interacting with a thermal reservoir, see Fig.~\ref{fig:sketch}.

The coupling with the bath at temperature $T$ determines the equilibrium state $\rho_\beta$ of the system where the subscript $\beta = 1/kT$ denotes inverse temperature and where $k$ is Boltzmann's constant. A discrete quantum operation $\Phi$ acting on the system will be compatible with the environment if it preserves $\rho_\beta$. We assume, moreover, that the map arises from a continuous time evolution starting at the identity and that it satisfies the  quantum detailed balance condition~\cite{agar}. These assumptions define the \emph{Davies maps}~\cite{davies} investigated in this work.

The paper is organized as follows. In Section~II we review some basic properties of quantum states and quantum maps, discuss the detailed balance condition and formally introduce the notion of Davies maps. In Section~III we provide a complete characterization of the one-qubit Davies maps, discuss possible uses of these maps in quantum optics and study their minimum output entropy. The case of Davies maps for three level systems is treated in Section~IV.

\section{Definitions and conventions}

\subsection{Qudits and quantum maps}

The states of an $N$-level quantum system are described by $N \times N$ density matrices $\rho$, these are Hermitian, positive semi-definite and have trace one
\begin{equation*}
\rho = \rho^\dagger,\enskip \rho \ge 0,\enskip \text{and } \tr\rho = 1.
\end{equation*}
States model the statistical data obtained by measuring  observables $A$, in case of Hermitian matrices of dimension $N$
\begin{equation*}
\< f(A) \> = \tr \bigl( \rho\, f(A) \bigr).
\end{equation*}
Furtheron $\c M_N$ will denote the algebra of $N \times N$ matrices with complex entries.

A quantum operation or quantum channel $\Phi$ is an affine transformation of the density matrices 
\begin{equation}
\rho \mapsto \rho' = \Phi(\rho)
\label{map}
\end{equation}
which is completely positive (CP). Remark that a quantum operation can be uniquely extended to a trace preserving linear transformation of the $N \times N$ matrices. Complete positivity means that the trivial extension $\Phi \otimes \id_d$ to a space of any dimension $d$ preserves positivity. The state transformation rule~(\ref{map}) is the Schr\"odinger picture of the action of the channel. One can use the corresponding Heisenberg picture as well, the map will still be completely positive but trace preserving becomes unity preserving. 

It was shown by Kraus~\cite{kraus} that a linear map $\Phi$ is completely positive if and only if it is of the form
\begin{equation}
\Phi(\rho) = \sum_i K_i \rho K_i^\dagger.
\label{kraus}
\end{equation}
If $\Phi$ is trace preserving, i.e.\ if $\Phi$ is a quantum operation, then the Kraus operators $K_i$ satisfy the relation 
\begin{equation*}
\sum_i K_i^\dagger K_i = \idty.
\end{equation*}

There is a useful condition that guarantees that a linear transformation $\Phi$ of the $N \times N$ matrices is completely positive. It suffices to check whether the map $\Phi \otimes \id_N$ transforms a maximally entangled state
\begin{equation}
|\Psi^+\> = \frac{1}{\sqrt N} \sum_{i=1}^N |i\> \otimes |i\>
\label{entanglement}
\end{equation}
into a positive matrix~\cite{choi,jam}. For a quantum operation $\Phi$
\begin{equation}
\bigl(\Phi \otimes \id_N\bigr)(\bigl|\Psi^+\> \<\Psi^+|\bigr)
= \frac{1}{N}\, \sum_{i,j=1}^N \Phi\bigl(|i\>\<j|\bigr) \otimes |i\>\<j|
\ge 0
\label{choi}
\end{equation}
is a density matrix called the Jamio\l kowski state or Choi matrix of $\Phi$.

A quantum map $\Phi$ is often represented in superoperator form $\Phi_S$. In this parametrization a density matrix $\rho$ is written as a vector $\vec\rho$ by ordering its elements in a column. Equation~(\ref{map}) can now be written as the action of a matrix $\Phi_S$ on $\vec\rho$
\begin{equation}
\vec{\rho'} = \Phi_S \vec\rho.
\label{supform}
\end{equation}

For a single qubit one can take the basis consisting in the identity and the three Pauli matrices $\vec\sigma=\{\sigma_{1},\sigma_{2},\sigma_{3}\}=\{\sigma_{x},\sigma_{y},\sigma_{z}\}$. A qubit state can then uniquely be written in terms of a vector in the closed  centred unit ball in three dimensional real space, called Bloch ball
\begin{equation*}
\rho = \tfrac{1}{2}\, (\id + \vec x \cdot \vec \sigma),\enskip
\vec x \in \Rl^3,\enskip \norm{\vec x} \le 1.
\end{equation*}
The quantum operation $\Phi$ in this basis will be called a quantum operation in the Bloch form $\Phi_B$. As $\Phi$ is affine, it transforms the Bloch ball into an ellipsoid inside the ball. Up to two orthogonal rotations of the basis the Bloch form of a one-qubit operation can be written as
\begin{equation*}
\Phi_B
= \begin{pmatrix}
1 &0 &0 &0 \\ \
\kappa_1 &\eta_1 &0 &0 \\
\kappa_2 &0 &\eta_2 &0 \\
\kappa_3 &0 &0 &\eta_3
\end{pmatrix}.
\end{equation*}
The vector $\vec\kappa$ describes the translation of the centre of the ellipsoid with respect to the centre of the Bloch ball and is zero for a bistochastic map. The absolute values of the $\eta_i$ are the lengths of the axes of the ellipsoid.

\subsection{Quantum Markov process}

The reduced evolution of a quantum system coupled to a general environment is hard to describe. Some additional assumptions make the problem more tractable. For example Markovianity means that the system interacts so weakly with the environment that it preserves its identity and that the evolution of the system does not depend on a whole history but only on its present state. A Markovian evolution in Schr\"odinger picture is described by a differential equation and an initial condition
\begin{equation}
\frac{\r d}{\r dt}\rho = \c G\rho
\enskip\text{and}\enskip
\rho(0) = \rho_0
\label{lindblad1}
\end{equation}
where $\c G$ is called the generator. It consists of a unitary part $\delta$ and a dissipative part $\c L$
\begin{equation}
\c G = i\delta + \c L.
\label{markovgenerator}
\end{equation}
Here 
\begin{equation}
\delta: \rho \mapsto [\rho \,,\, H]
\label{lindblad3}
\end{equation} 
is a commutator with an Hermitian system Hamiltonian $H$ and $\c L$ is a Lindblad operator
\begin{equation}
\c L: \rho \mapsto \sum_\alpha \Bigl( K_\alpha\, \rho\, K_\alpha^\dagger - \tfrac{1}{2}\, \bigl\{ K_\alpha^\dagger K_\alpha \,,\, \rho \bigr\} \Bigr).
\label{lindblad2}
\end{equation}
The $K_\alpha \in \c M_N$ can be compared with the Kraus operators~(\ref{kraus}). The splitting~(\ref{markovgenerator}) into a unitary and a dissipative part is not unique. The solution of equation~(\ref{lindblad1}) is
\begin{equation*}
\rho(t) = \r e^{\c G t} \rho_0.
\end{equation*}
The form of a Lindblad operator~(\ref{lindblad2}) guarantees that for any time $t \ge 0$ the map $\exp(\c G t)$ is completely positive, as was shown by Gorrini, Kossakowski, Sudarshan~\cite{gorini} and Lindblad~\cite{lindblad}.

A linear transformation $\c G$ on the $N \times N$ matrices is a generator of a semi-group of CP maps if and only if it satisfies a criterion similar to the Choi-Jamio\l kowski criterion~(\ref{choi}) for CP maps. For $\c G$ that kill the trace, $\tr \c G(A) = 0$ for all $A$, one has to check the positivity of the Choi matrix of the generator on the subspace perpendicular to the maximally entangled vector~(\ref{entanglement}). Such a $\c G$ is then decomposable in the form~(\ref{lindblad2}). This type of condition is called conditional complete positivity~\cite{wolf}.

\subsection{Quantum detailed balance condition}
\label{s1.3}

Our model is an $N$-level quantum system weakly coupled to a thermal bath at temperature $T$. The quantum system is described by its effective Hamiltonian $H$ which is diagonal with eigenvalues $\{e_i\}$. The coupling with the bath distinguishes the equilibrium state $\rho_\beta$ at inverse temperature $\beta = 1/kT$, where $k$ is Boltzmann's constant. The density matrix $\rho_\beta$ is called a Gibbs state, it is also diagonal in the eigenbasis of the Hamiltonian
\begin{equation*}
\rho_\beta = \frac{1}{\c Z}\, \exp(-\beta H)
\enskip\text{or}\enskip
(\rho_\beta)_{ij} = \frac{1}{\c Z}\, \r e^{-\beta e_i}\, \delta_{ij}.
\end{equation*}
Here $\c Z$ is a normalization factor called partition function
\begin{equation*}
\c Z = \sum_i \r e^{-\beta e_i}.
\end{equation*}

The weak-coupling limit yields semi-groups in continuous time of completely positive maps which enjoy the \emph{quantum detailed balance} property, for a detailed treatment see~\cite{davies,alicki}. Such semi-groups are called Davies semi-groups and they are to be compared with classical Glauber dynamics. The generator of a Davies semi-group splits into a unitary and a dissipative part as in~(\ref{markovgenerator}) in such a way that the unitary and the dissipative parts decouple. Moreover, the dissipative part causes jumps between the eigenstates of the system Hamiltonian with transition rates that obey the micro-reversibility condition: the particle flow in equilibrium from level $i$ to level $j$ is exactly balanced by the reverse flow. Furthermore, off-diagonal elements of a density matrix decay at rates that are suitably adapted to the diagonal transition rates. A more precise definition follows below. 

A Davies map is an element of a Davies semi-group at some particular time. Due to the detailed balance property of the generator, a Davies map will satisfy the quantum detailed balance property with respect to the equilibrium state $\rho_\beta$~\cite{agar}. This property is stronger than simple invariance and is usually expressed in terms of a self-adjointness condition with respect to the scalar product defined by the equilibrium state. The converse is, however, not true: there are quantum detailed balance maps that don't sit on a Davies semi-group, for some general results on the characterizing maps that sit on semi-groups see~\cite{wolf}. 

A Davies map is defined as follows
\begin{enumerate}
\item
$\Phi = \exp(\c Gt)$,
\item
$\c G = i\delta + \c L$ with $\delta$ and $\c L$ as in~(\ref{lindblad3}) and (\ref{lindblad2}),
\item
$\delta$ and $\c L$ commute, and
\item
$\c L$ is self-adjoint with respect to the Gibbs state defined by the unitary dynamics of the system in the following sense, for all $A,B \in \c M_N$
\begin{equation}
\tr \c L \bigl( B\, \rho_\beta \bigr) \, A = \tr \c L \bigl( \rho_\beta\, A \bigr) \, B. 
\label{herm}
\end{equation}
\end{enumerate}

Let us consider the matrix representation of the dissipative part $\c L$. The eigenvectors of the derivation $\delta(\cdot) = [H \,,\, \cdot]$ are the elementary matrix units $|i\> \<j|$ with corresponding eigenvalues $e_i - e_j$. The third property from the above list guarantees that $\c L$ has the same eigenspaces as $\delta$. Therefore the diagonal elements of a density matrix are not mixed with off-diagonal ones during the evolution determined by  $\c L$. Furthermore, if the energy differences in the Hamiltonian are not degenerated off-diagonal elements are not mixed among themselves. Therefore the superoperator matrix $\c L_S$ is block diagonal in the decomposition
\begin{equation*}
\c M_N = \r{Span}\bigl( \{|i\> \<i| \mid i= 1, 2, \ldots N\}\bigr) \loplus \Bigl( \loplus_{i \neq j} \Cx |i\> \<j| \Bigr).
\end{equation*}
The fourth property guarantees that the restriction of $\c L$ to the diagonal matrices is a classical detailed balance generator with respect to the probability vector of the eigenvalues of the Gibbs state and that the eigenvalues of $\c L$ corresponding to non-diagonal matrix elements are real. These conclusions also hold true for a quantum operation $\Phi = \exp(\c Gt)$ and therefore for a Davies map. The requirement that $\c G$ and hence $\Phi$ should be completely positive will be imposed by checking the positivity of the Choi matrix of the generator on the subspace perpendicular to the maximally entanglement state. The unitary evolution generated by $\delta$ will play no role in what follows and we therefore simply omit it. It follows also from the fourth condition that the equilibrium state $\rho_\beta$ is invariant. 
\begin{enumerate}
\item[5.]
Often one imposes ergodicity which requires that $\rho_\beta$ is the unique invariant state: for any density matrix $\rho$
\begin{equation*}
\lim_{t \to \infty} \exp(\c Gt)(\rho) = \lim_{t \to \infty} \exp(\c Lt)(\rho) = \rho_\beta
\end{equation*}
\end{enumerate}

Consider the following scalar product on the observables
\begin{equation*}
\< X \,,\, Y \>_\beta := \tr \bigl( \rho_\beta^{-1} X^\dagger Y \bigr).
\end{equation*}
The detailed balance relation~(\ref{herm}) for the dissipative part of the generator can be rewritten as
\begin{equation*}
\< X \,,\, \c L(Y) \>_\beta = \< \c L(X) \,,\, Y \>_\beta.
\end{equation*}
Hence $\c L$ is Hermitian for this scalar product and the same is true for $\exp(\c Lt)$. As a consequence both $\c L$ and $\exp(\c Lt)$ have a real spectrum.

\section{Davies maps on a two-level system}

\subsection{One-qubit maps}

We write the Hamiltonian of the system in its eigenbasis
\begin{equation*}
H
= \begin{pmatrix}
e_1 &0 \\
0 &e_2
\end{pmatrix} \enskip\text{with}\enskip e_1 > e_2.
\end{equation*}
This implies that the Gibbs state consisting of elements of a probability vector $(p,1-p)$ is of the form
\begin{equation}
\rho_\beta
= \begin{pmatrix}
p &0 \\
0 &1-p
\end{pmatrix}
= \frac{1}{\c Z}\, \begin{pmatrix}
\exp(-\beta e_1) &0 \\
0 &\exp(-\beta e_2)
\end{pmatrix},
\label{qubitgibbs}
\end{equation}
where $\beta = 1/kT$ and the partition function reads
\begin{equation*}
\c Z := \exp(-\beta e_1) + \exp(-\beta e_2).
\end{equation*}

The general superoperator form of a qubit Davies map is 
\begin{equation}
\Phi_S
= \begin{pmatrix}
1-a &0 &0 &a\,p/(1-p) \\
0 &c &0 &0 \\
0 &0 &c &0 \\
a &0 &0 &1 - a\,p/(1-p)
\end{pmatrix},
\label{davies}
\end{equation}
where $a$ and $c$ are parameters of the map, while $(p,1-p)$ is the vector of diagonal entries of the stationary state~(\ref{qubitgibbs}). At infinite temperature $\beta = 0$ the map is bistochastic, i.e.\ the uniform state is stationary. We still need to impose the condition that $\Phi_S$ lies on a one-qubit Davies semi-group of completely positive maps. This provides a full characterization of the allowed values of the parameters $a$ and $c$,
\begin{equation}
0 \le a \le 1, \enskip a + p \le 1, \enskip\text{and}\enskip 0 \le c \le \sqrt{1 - a/(1-p)}.
\label{davies2}
\end{equation}

Because of the relation $\Phi_S = \exp(\c Lt)$ the parameters $a$ and $c$ change in time according to
\begin{equation*}
a = (1-p) \bigl( 1-\exp(-At) \bigr) \enskip\text{and}\enskip c = \exp(-\Gamma t),
\end{equation*}
where $A$ and $\Gamma$ are constants such that
\begin{equation*}
\Gamma \ge \tfrac{1}{2}\, A \ge 0.
\end{equation*}
This constraint on the decay rate of off-diagonal elements with respect to the rate at which equilibrium is attained has been known for a long time, see e.g.~\cite{alicki,benatti}. The constants $A$, $\Gamma$ and the stationary vector~(\ref{qubitgibbs}) determine a generator $\c L$ of a semi-group $\{ \exp(\c L t) \mid t \ge 0 \}$.

\begin{figure}[h]
\centering
\scalebox{.4}{\includegraphics{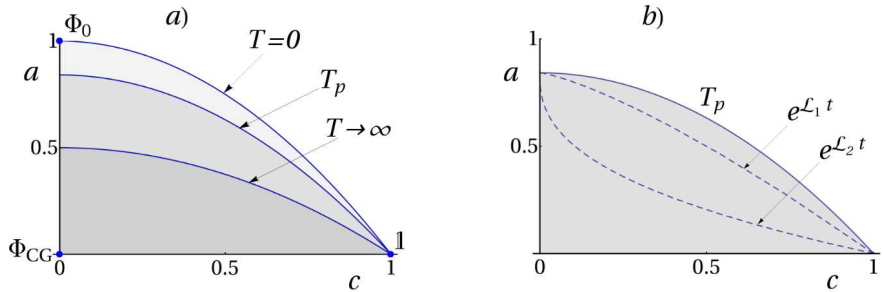}}
\caption{Panel a) Allowed range of parameters in the $(c,a)$ plane for one-qubit Davies maps~(\ref{davies}) shown for three different temperatures: $T=0$, $T_p \in (0,\infty)$, and $T \rightarrow \infty$. The corners of the plot denote the identity channel $\idty$, coarse graining $\Phi_{\r{CG}}$ and a decaying channel $\Phi_0: \rho \rightarrow |0\> \<0|$. 
Panel b) The dashed lines in the allowed region at temperature $T_p$ represent the continuous time evolution $\exp(\c L t)$ for two typical generators $\c L_1$ and $\c L_2$.}
\label{fig:ryzunki}
\end{figure}

\begin{figure}[h]
\centering
\scalebox{.6}{\includegraphics{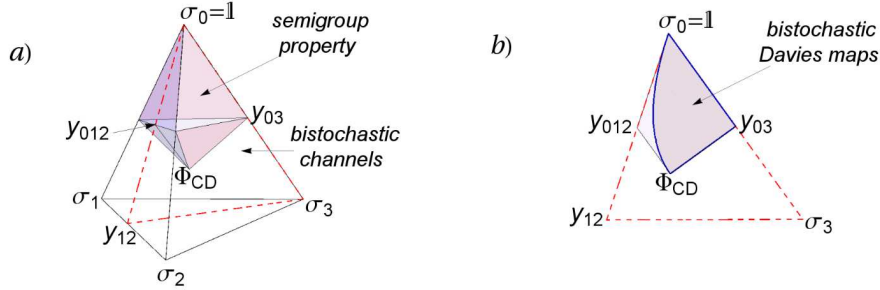}}
\caption{a) The tetrahedron of one-qubit bistochastic maps contains the shaded region $P$ of maps of the form $\exp(\c L t)$. b) The bistochastic Davies maps belong to the cross section of the set $P$ with the plane determined by $\sigma_0 = \idty$, $\sigma_3 = \sigma_z$ and the completely depolarizing channel $\Phi_{\r{CD}}: \rho \mapsto \frac{1}{2}\, \idty$. Its boundary is given by~(\ref{davies2}).}
\label{fig:bisto}
\end{figure}

The allowed values of the parameters $a$ and $c$ occupy the region shown in Fig.~\ref{fig:ryzunki} which depends on temperature. For higher temperatures the allowed region shrinks. In the limiting case $T\to \infty$ the equilibrium state is maximally mixed, $\rho_{\beta=0} = \frac{1}{2}\,\idty$. This case corresponds to unital and thus bistochastic maps. The set of all one-qubit bistochastic maps forms a regular tetrahedron spanned by the three Pauli channels $\sigma_1$, $\sigma_2$, and $\sigma_3$ and by the identity channel $\sigma_0$. The set of bistochastic maps lying on a semi-group is a proper subset: two tetrahedra glued at their base and localized in the vicinity of the identity map. The bistochastic Davies maps form a two-dimensional sub-region, see Fig.~\ref{fig:bisto}.

\begin{figure}[ht]
\centering
\scalebox{.3}{\includegraphics{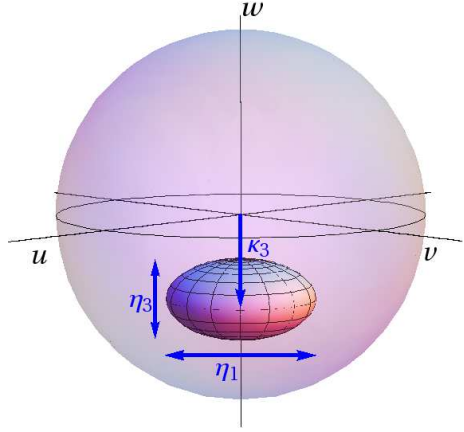}}
\caption{The Bloch ball is transformed by the Davies map~(\ref{qubitbloch}) into an ellipsoid centred at $\vec x = (0,0,w_{\r{eq}} := 2p - 1)$ with axes $(\eta_1,\eta_1,\eta_3)$.}
\label{fig:detbal}
\end{figure}

In Bloch parametrization a qubit Davies map takes the form
\begin{equation}
\Phi
= \begin{pmatrix}
1 &0 &0 &0 \\
0 &\eta_1 &0 &0 \\
0 &0 &\eta_2 &0 \\
\kappa_3 &0 &0 &\eta_3
\end{pmatrix},
\label{qubitbloch}
\end{equation}
where $\eta_1 = \eta_ 2 = c$, $\eta_3 = 1 - a/(1-p)$, and $\kappa_3 = a(2p - 1)/(1-p)$. Such a map acts on the Bloch ball as shown in Fig.~\ref{fig:detbal}.

\subsection{Davies maps in quantum optics}

Quantum dynamics consisting of a unitary part and a dissipative part are often considered in optics, the unitary evolution is caused by a driving external field while the dissipation is a consequence of a coupling to an external reservoir. The simplest instance is that of a two level atom. Its states are determined by a three component Bloch vector $\vec{x}=(u,v,w)$. The component $w$, the difference between the diagonal elements of the density matrix, measures the inversion of populations and the off-diagonal elements $u$ and $v$ are associated with an atomic dipole operator, $u$ in phase and $v$ out of phase with respect to the external field. The dissipation due to the interaction with the environment decreases the length of the Bloch vector and therefore drives a pure initial state towards a more mixed state. 

The total evolution which is a composition of a unitary and a dissipative part is described by the Bloch equation, this is an affine differential equation for the Bloch vector. Affinity is a consequence of the linearity of quantum mechanics: an initial mixture of pure states evolves to the same mixture of evolved pure states. A few empirical constants $\tau_u$, $\tau_v$, $\tau_w$, $\Delta$, $\Omega$, and $w_{\r{eq}}$ enter the equations for $\vec{x}=(u,v,w)$
\begin{equation*}
\left\{\begin{aligned}
\frac{{\rm d}}{{\rm d}t} u &= -\frac{1}{\tau_u} - \Delta v \\
\frac{{\rm d}}{{\rm d}t} v &= -\frac{1}{\tau_v} + \Delta u + \Omega w \\
\frac{{\rm d}}{{\rm d}t} w &= -\frac{1}{\tau_w} (w - w_{\r{eq}}) - \Omega v\\
\end{aligned}\right.
\end{equation*}
The constants have a physical interpretation: $\tau_w$ is the rate at which inversion of the populations tends to the equilibrium state, $\tau_u$ and $\tau_v$ are decay rates of the atomic dipole, $\Delta$ is the field-atom detuning, and $\Omega$ is the Rabi frequency of oscillation between the ground and excited state of the atom. The parameter $w_{\r{eq}}$ fixes the equilibrium inversion.

A semi-group of completely positive maps yields analogous equations for the Bloch vector. The generator $\c L_B$ of a Davies map in Bloch parametrization is a 4 dimensional matrix acting on the vector $(1,u,v,w)^{\s T}$ where $(u,v,w)$ is the Bloch vector. We use the notation $\tau_1 =\tau_2= \tau_u = \tau_v$ and $\tau_3 = \tau_w$. The relaxation times $\tau_1$ and $\tau_3$ and the equilibrium inversion $w_{\r{eq}}$ determined by an energy difference and a temperature are physical parameters while $\eta_1$, $\eta_2$, and $\kappa_3$ describing the ellipsoid obtained as the image of the pure qubit states is rather a geometrical parametrization
\begin{equation}
\c L_B
= \begin{pmatrix}
0 &0 &0 &0 \\
0 &-1/\tau_1 &0 &0 \\
0 &0 &-1/\tau_1 &0 \\
w_{\r{eq}}/\tau_3 &0 &0 &-1/\tau_3
\end{pmatrix}.
\label{generator}
\end{equation}
The variable $w_{\r{eq}}$ is related to the equilibrium state $\rho_\beta = \begin{pmatrix} p &0 \\ 0 &1-p \end{pmatrix}$ by $w_{\r{eq}} = 2p - 1$. The Bloch equations describing a Davies map are:
\begin{equation*}
\left\{\begin{aligned}
\frac{{\rm d}}{{\rm d}t} u &= -\frac{1}{\tau_1} \\
\frac{{\rm d}}{{\rm d}t} v &= -\frac{1}{\tau_1} \\
\frac{{\rm d}}{{\rm d}t} w &= -\frac{1}{\tau_3} (w - w_{\r{eq}}).
\end{aligned}\right.
\end{equation*}
By changing the basis one obtains the generator matrix in superoperator form
\begin{equation*}
\c L_S =
\begin{pmatrix}
-(1-p)/\tau_3 &0 &0 &p/\tau_3 \\
0 &-1/\tau_1 &0 &0 \\
0 &0 &-1/\tau_1 &0 \\
(1-p)/\tau_3 &0 &0 &-p/\tau_3
\end{pmatrix}
\end{equation*}
It can be checked that the outer (1,4)-block of this matrix satisfies the classical detailed balance condition with respect to the invariant probability vector $(p,1-p)$.

The quantum Davies map $\exp(\c Lt)$ reads in Bloch parametrization
\begin{equation}
\exp(\c L_B t)=\begin{pmatrix}
1 &0 &0 &0 \\
0 &\r e^{-t/\tau_1} &0 &0 \\
0 &0 &\r e^{-t/\tau_1} &0 \\
(1 - \r e^{-t/\tau_3}) w_{\r{eq}} &0 &0 &\r e^{-t/\tau_3}
\end{pmatrix}.
\label{blochdecay}
\end{equation}
The diagonal entries of this matrix belong to the spectrum of $\exp(\c Lt)$ which is real. In superoperator form~(\ref{supform}) the same map is represented by the matrix 
\begin{equation*}
\exp(\c L_S t)=\begin{pmatrix}
1 - (1 - \r e^{-t/\tau_3})(1 - p) &0 &0 &(1 - \r e^{-t/\tau_3})\, p \\
0 &\r e^{-t/\tau_1} &0 &0 \\
0 &0 &\r e^{-t/\tau_1} &0 \\
(1 - \r e^{-t/\tau_3})(1 - p) &0 &0 &1 - (1 - \r e^{-t/\tau_3})\,p
\end{pmatrix}.
\end{equation*}
The conditions for lying on a semi-group of completely positive maps are
\begin{equation}
\tau_1=\tau_2 \ge 0,\enskip \tau_3 \ge 0,\enskip \text{and } \tau_1 \le 2\,\tau_3.
\label{wynik}
\end{equation}
These conditions are known as necessary conditions for a master equation of Lindblad type~\cite{kimura}. In this article the author obtains also a map of the form~(\ref{generator}) and the relations~(\ref{wynik}). In~\cite{daffer} the map~(\ref{blochdecay}) was considered in connection with the squeezed vacuum channel. It was derived directly from a specific form of the Lindblad generator.

\subsection{Minimal output entropy}

There are several possibilities for quantifying the randomizing properties of a channel both classical and quantum. We focus here on the minimal von Neumann output entropy~\cite{shor, kingruskai} but Renyi entropies~\cite{giovanetti, werner} are also considered. A different approach, the map entropy~\cite{ZB04, ziman}, uses the Jamio\l kowski state and, at least for classical channels, the entropy of an associated Markov process has also been investigated~\cite{roga}. 

The minimal output entropy is defined as
\begin{equation}
\s S_{\r{min}}(\Phi) := \min_\rho \s S(\Phi(\rho)),
\label{moe}
\end{equation}
where $\s S$ is the von Neumann entropy
\begin{equation*}
\s S(\rho) := -\tr (\rho\, \log\rho),\enskip \rho \text{ a density matrix.}
\end{equation*}
Because the entropy is concave we can always find a pure minimizer for~(\ref{moe}). In Fig.~\ref{fig:detbal} an inverse image of a point on the ellipsoid nearest to the Bloch sphere is such a pure minimizer. As was shown before the image of the Bloch ball under a Davies map has rotational symmetry with respect to an axis of the Bloch ball. Therefore, in this case, there is whole family of minimizers. In order to calculate the minimal output entropy it suffices to consider density matrices with all entries positive. This generalizes to higher dimensional problems.

In a number of cases the minimizer for the output entropy is known. The simplest situation is probably a classical $N$ state stochastic map. This is given in terms of a stochastic matrix: a $N \times N$ matrix with non-negative entries and column sums equal to one. Each column of such a matrix is a probability vector with a corresponding Shannon entropy. The minimal output entropy is easily seen to be the smallest of these entropies, in other words, the minimizer is the degenerate measure corresponding to the minimizing column. Another well-known case is that of Bosonic Gaussian channels. It was conjectured for a long time that the minimizer is a Gaussian state. This conjecture was recently proven in~\cite{lloyd}. Also for Davies maps there are natural candidates for a minimizer: the eigenstates of the system Hamiltonian. This turns out to be false even in the qubit case, at least for maps that are sufficiently close to the identity. After a sufficiently long time the Davies map will generically be close to the projection on the equilibrium state and therefore behave more or less like a classical stochastic map. One indeed observes that in such a situation the minimizer is an eigenstate of the Hamiltonian.

Let us take the input state
\begin{equation*}
\rho
= \begin{pmatrix}
\mu &\nu \\
\nu &1 - \mu
\end{pmatrix},\enskip 0 \le \mu \le 1, \text{ and } 0 \le \nu \le \sqrt{\mu(1-\mu)},
\end{equation*}
which is pure if and only if $\nu^2 = \mu(1-\mu)$. After acting on this state by the Davies map
\begin{equation*}
\Phi_S
= \begin{pmatrix}
1 - a &0 &0 &b \\
0 &c &0 &0 \\
0 &0 &c &0 \\
a &0 &0 &1 - b
\end{pmatrix},
\end{equation*}
where $b = ap/(1-p)$ like in~(\ref{davies}) and~(\ref{davies2}), one gets
\begin{equation*}
\rho' 
= \begin{pmatrix}
(1-a)\mu + b(1-\mu) &c\nu \\
c\nu &a\mu + (1-b)(1-\mu)
\end{pmatrix}.
\end{equation*}
The eigenvalues $\xi_{\pm}$ of $\rho'$ are
\begin{equation*}
\xi_{\pm} \ = \ \frac{1}{2}\, \bigl( 1 \pm \sqrt{1 - 2b + 2(-1+a+b)\mu} + 4c^2\nu^2 \bigr).
\end{equation*}
The entropy reaches its minimum when the difference between the two eigenvalues is maximal. Therefore we need to find the maximum of the expression under square root substituting $\nu^2$ by $\mu(1-\mu)$. The second derivative of this expression is equal to $8(1-a-b)^2 - c^2$. A number cases have to be distinguished:
\begin{enumerate}
\item
If $c^2 \le (1-a-b)^2$ then the minimizer is $\mu = 0$ which is the ground state of the system Hamiltonian and the corresponding output state is diagonal
\begin{equation}
\Phi(\rho_{\r{min}})
= \begin{pmatrix} 1-b &0 \\ 0&b \end{pmatrix}.
\label{min1}
\end{equation}
\item
If $(1-a-b)^2 \le c^2 \le (1-a-b)(1-2b)$ the minimizer is again $\mu = 0$. In this case the function attains its maximum outside the set of admissible values of $\mu$ and therefore $\mu$ is an endpoint of $(0,1)$.
\item
If $(1-a-b)(1-2b) \le c^2$ the the minimizer is determined by
\begin{equation*}
\mu_0 = \frac{(a+b-1)(2b-1) - c^2}{2(a+b-1)^2 - 2c^2}.
\end{equation*}
In this case it is a true superposition of ground and exited states of the system Hamiltonian. The eigenvalues of the output state are
\begin{equation}
\xi_{\pm} \ = \ 
\frac{1}{2}\, \Bigl( 1 \pm \sqrt{\frac{c^2 (c^2 + a(2-4 b) + 2b-1)}{c^2 - (-a-b+1)^2}} \Bigr).
\label{min2}
\end{equation}
\end{enumerate}

The entropy is easily computed from the formulas~(\ref{min1}) and~(\ref{min2}). The condition $c^2 \le (1-a-b)^2$ written in terms of decay constants translates to $\tau_1 \le \tau_3$.

\section{Davies maps on a three-level system}

As discussed at the end of Section~\ref{s1.3}, a generic Davies map $\Phi$ stochastically mixes the diagonal components of a density matrix and scales down the off-diagonal entries. The full Davies superoperator acting on a three-level quantum system
is therefore represented by a matrix
\begin{equation}
\Phi_S
= \begin{pmatrix}
1 - F_{21} - F_{31} &0 &0 &0 &F_{12} &0 &0 &0 &F_{13} \\
0 &\lambda_3 &0 &0 &0 &0 &0 &0 &0 \\
0 &0 &\lambda_2 &0 &0 &0 &0 &0 &0 \\
0 &0 &0 &\lambda_3 &0 &0 &0 &0 &0 \\
F_{21} &0 &0 &0 &1 - F_{12} - F_{32} &0 &0 &0 &F_{23} \\
0 &0 &0 &0 &0 &\lambda_1 &0 &0 &0 \\
0 &0 &0 &0 &0 &0 &\lambda_2 &0 &0 \\
0 &0 &0 &0 &0 &0 &0 &\lambda_1 &0 \\
F_{31} &0 &0 &0 &F_{32} &0 &0 &0 &1 - F_{13} - F_{23}
\end{pmatrix}.
\label{duza}
\end{equation}

The stochastic matrix
\begin{equation}
F
= \begin{pmatrix}
1 - F_{21} - F_{31} &F_{12} &F_{13} \\
F_{21} &1 - F_{12} - F_{32} &F_{23} \\
F_{31} &F_{32} &1 - F_{13} - F_{23}
\end{pmatrix}
\label{klasyczna}
\end{equation}
satisfies the classical detailed balance condition with respect to the probability vector $(p_1,p_2,p_3)$ where the $p_i$ are the eigenvalues of $\rho_\beta$, determined by the energy levels of the system Hamiltonian and by the temperature
\begin{equation}
F_{ij}\, p_j = F_{ji}\, p_i,\enskip i,j = 1,2,3.
\label{cdb3}
\end{equation}
As in the qubit case we assume that the eigenvalues of the system Hamiltonian are strictly ordered: $e_1 > e_2 > e_3$, put differently $p_1 < p_2 < p_3$. Assuming that all entries of $F$ are strictly positive, the conditions~(\ref{cdb3}) are satisfiable if and only if
\begin{equation*}
F_{12} < F_{21},\enskip F_{13} < F_{31},\enskip F_{23} < F_{32},\enskip \text{and}\enskip F_{12}\, F_{23}\, F_{31} = F_{13}\, F_{32}\, F_{21},
\end{equation*}
one of the inequalities being redundant. The stationary vector is given by
\begin{equation*}
p_1 = \frac{F_{12} F_{13}}{F_{12} F_{13} + F_{12} F_{31} + F_{13} F_{21}}
\end{equation*}
and the other components $p_2$ and $p_3$ are obtained by cyclically permuting this expression.

Another way to look at the stochastic map $F$ is to say that it is a special case of a quantum map: it mixes the diagonal elements of a density matrix according to $F$ and  destroys the off-diagonals. Complete positivity is automatically satisfied because the Choi matrix of this map is diagonal with the entries of $F$ on the diagonal.

Furthermore, if $\Phi$ is a Davies map then it lies on a Davies semi-group and this implies that $F$ lies on a classical detailed balance semi-group of stochastic maps
\begin{equation*}
F = \exp(G) %\enskip\text{with}\enskip G \text{ a classical detailed balance generator, and $t$ is set to unity}.
\end{equation*}
with $G$ a classical detailed balance generator and $t=1$. In particular this means that
\begin{equation*}
G
= \begin{pmatrix}
-G_{21} - G_{31} &G_{12} &G_{13} \\
G_{21} &-G_{12} - G_{32} &G_{23} \\
G_{31} &G_{32} &-G_{13} - G_{23}
\end{pmatrix}
\end{equation*}
with
\begin{equation*}
G_{ij}\, p_j = G_{ji}\, p_i,\enskip i,j = 1,2,3.
\end{equation*}

As argued at the end of Section~\ref{s1.3} detailed balance generators have real eigenvalues. A property which we have to check in order to ensure the existence of a semi-group $F = \exp(G)$ is positivity of the eigenvalues of $F$ because only in this case there exist a real logarithm of $F$. The three eigenvalues of $F$ in decreasing order are
\begin{equation}
\sigma(F)=\{1,A+B,A-B\},
\label{eig}
\end{equation}
where
\begin{equation*}
A = \tfrac{1}{2}\, (\tr F - 1)
\enskip\text{and}\enskip
B = \tfrac{1}{2}\, \sqrt{2 \tr F^2 -(\tr F)^2 + 2\tr F - 3}.
\end{equation*}
The existence of a logarithm therefore requires the positivity conditions
\begin{equation*}
0 \le A \enskip\text{and}\enskip B \le A,
\end{equation*}
these are equivalent with
\begin{equation*}
1 \le \tr F \enskip\text{and}\enskip \tr F^2 + 2 \tr F \le 2 + \bigl( \tr F \bigr)^2.
\end{equation*}

The last criterion which has to be satisfied is that $G = \log(F)$ is a generator which means that its off-diagonal elements must be positive. It is possible to compute $\log(F)$ straightforwardly, see the appendix. The last condition that guarantees that $F$ belongs to a semi-group reads
\begin{equation}
G_{12} \ge 0,\enskip G_{23} \ge 0,\enskip \text{and}\enskip G_{31} \ge 0.
\label{nierownosci}
\end{equation}
These conditions are rather complicated. There are three of them relating the spectrum~(\ref{eig}) of $F$ with its entries. We just write down a single one explicitly derived in the appendix:
\begin{equation}
y_1 (1 - A + B) \log(A + B) \ \le \  y_2  (1 - A - B) \log(A - B)
\label{G}
\end{equation}
where
\begin{align*}
y_1 
&:= 2B - F_{12} - F_{21} + F_{13} - F_{31} + F_{23} - F_{32} + 2 \frac{F_{23} F_{31}}{F_{21}} \enskip\text{and} \\
y_2 
& := 4 - 2B - F_{12} - F_{21} + F_{13} - F_{31} + F_{23} - F_{32} + 2 \frac{F_{23} F_{31}}{F_{21}}.
\end{align*}
The other two conditions are obtained by cyclically permuting that of above.

We now return to the full quantum map $\Phi$. Clearly the $\lambda_j$ in~(\ref{duza}) must be positive as they have to be exponentials of real numbers. We write $r_j = - \log\lambda_j$. An additional condition should be imposed in the quantum case. The generator of $\Phi$ has to be conditionally complete positive which means that
\begin{equation*}
- \begin{pmatrix}
G_{21} + G_{31} &r_3 &r_2 \\
r_3 &G_{12} + G_{32} &r_1 \\
r_2 &r_1 &G_{13} + G_{23}
\end{pmatrix} \ge 0 \enskip\text{on}\enskip \bigl( \{1,1,1\}^{\s T} \bigr)^\perp.
\end{equation*}
This translates into the conditions
\begin{equation*}
r_1 + r_2 + r_3 \ge G_{12} + G_{23} + G_{31} + G_{21} + G_{32} + G_{13} 
\end{equation*}
and
\begin{equation*}
\begin{split}
&r_1^2 + r_2^2 + r_3^2 + 2\bigl( r_1 (G_{21} + G_{31}) + r_2 (G_{32} + G_{12}) + r_3 (G_{31} + G_{12}) \bigr) \\
&\quad\le 2\bigl( r_1 r_2 + r_2 r_3 + r_3 r_1 \bigr) + \bigl( (G_{21} + G_{31}) (G_{32} + G_{12}) \\
&\phantom{\quad\le\ }+ (G_{32} + G_{12}) (G_{13} + G_{23}) + (G_{13} + G_{23}) (G_{21} + G_{31}) \bigr)
\end{split}
\end{equation*}

Together with~(\ref{nierownosci}) the above relations guarantee that $\Phi$ is a Davies map. We assumed that the Bohr spectrum of the system is non-degenerate, that is that there are no degeneracies in the non-zero differences of energy levels of the Hamiltonian. If it is not the case then blocks appear on the diagonal of~(\ref{duza}).

\section{Conclusion}

We studied a class of quantum operations called Davies maps. They satisfy the quantum detailed balance condition and the semi-group property $\Phi = \exp(\c Gt)$. Moreover, the unitary and the dissipative parts of the generator $\c G$ commute. This property implies that the dynamics of populations, the diagonal elements of a density matrix $\rho$, separates from the evolution of coherences, the off-diagonal elements of $\rho$.

For a single qubit the complete characterization of the set of Davies maps has been worked out. This three dimensional set of maps is determined by parameters $a$ and $c$ subject to the conditions~(\ref{davies2}) at a given temperature $T$. These parameters determine the lengths of the axes of the ellipsoid obtained by applying the Davies map $\Phi$ to the Bloch sphere. Thus they are related to both relaxation rates $\tau_1 = \tau_2$ and $\tau_3$ often used to describe a non-unitary one-qubit map in quantum optics. The constraints~(\ref{wynik}) which determine the set of Davies maps are found to be equivalent with conditions for a master equation of the Lindblad type.

Alternatively, one may parametrize the set of one-qubit Davies maps by the following three numbers: the vertical shift $\kappa_3$ of the centre of the Bloch ball, see Fig.~\ref{fig:detbal}, the volume of the image of the ball and its asymmetry parameter. It is easily seen that the first parameter depends of the temperature, while the two others are functions of $a$ and $c$.

So far we restricted our attention to the low dimensional cases $N=2$ and $N=3$ but some remarks can be made on general $N$-level Davies maps. Consider such a system described by a Hamiltonian with a non-degenerate spectrum. A Davies channel that satisfies all conditions
specified in Sec.~\ref{s1.3} is described by a block matrix $\Phi$ of size $N^2$. The outer block is a stochastic matrix $F$ which describes the transfer of populations. This matrix obeys the classical detailed balance condition, see~(\ref{duza}) and~(\ref{klasyczna}) written for $N=3$. The number of parameters needed to describe $F$ is $(N^2+N)/2 - 1$: the number of elements on the diagonal of $F$ and above it minus one due to the trace-preserving condition. There are also $(N^2-N)/2$ elements of $\Phi$ which describe the evolution of quantum coherences, the parameters $\lambda_i$ in~(\ref{duza}). Together this gives $d = N^2 - 1$ parameters necessary to specify a Davies map acting on an $N$-level system. This is exactly the number of parameters required to characterize a density matrix of this size. Thus the Davies maps forms a measure zero subset of all stochastic maps, a set of dimension $D = N^4 - N^2 = N^2 d$.

We computed the minimal output entropy of a one-qubit Davies map and found that in the quantum regime, close to the identity map, the minimizer is not equal to the obvious candidate: the eigenstate of the system Hamiltonian which remains as pure as possible under the stochastic mixing of diagonal elements. It is then interesting to ask for the additivity of minimal output entropy for this class of maps. Independently whether this property holds true, we believe that Davies maps form a physically relevant class of quantum operations, an interesting subject of further study.

\bigskip
\noindent
\textbf{Acknowledgements}

It is a pleasure to thank R. Alicki for helpful remarks. We acknowledge financial support by the bilateral project BIL05/11 between Flanders and Poland, by the Polish Ministry of Science and Higher Education under the grant the special grant number DFG-SFB/38/2007 and the Marie Curie Transfer of Knowledge Grant MTKD-CT-2004-517186 “Correlations in Complex Systems”(COCOS).

\appendix

\section{Derivation of inequality (\ref{G})}

Let $X$ be a $3 \times 3$ complex matrix and assume for simplicity that it has non-degenerate eigenvalues $\xi_1$, $\xi_2$, and $\xi_3$. This implies that $X$ is diagonalizable. Let $g$ be a complex function defined on a domain containing the eigenvalues of $X$, then
\begin{equation*}
g(X) = g_0 \idty + g_1 X + g_2 X^2
\end{equation*}
where $g_0$, $g_1$, and $g_2$ are complex functions that only depend on the eigenvalues of $X$. More explicitly
\begin{align*}
g_0
&= -\frac{\xi_2 \xi_3}{(\xi_1 - \xi_2) (\xi_3 - \xi_1)}\, g(\xi_1) + \text{cycl.\ perm.} \\
g_1
&= \frac{\xi_2 + \xi_3}{(\xi_1 - \xi_2) (\xi_3 - \xi_1)}\, g(\xi_1) + \text{cycl.\ perm.} \\
g_2
&= -\frac{1}{(\xi_1 - \xi_2) (\xi_3 - \xi_1)}\, g(\xi_1) + \text{cycl.\ perm.}
\end{align*}

We can use this to compute the logarithm of the stochastic matrix $F$ in~(\ref{klasyczna}). Using the parametrization~(\ref{eig}) for the eigenvalues of $F$ we obtain expression~(\ref{G}).

\end{document}